\RequirePackage{snapshot}
\documentclass[a4paper,12pt,oneside]{article}
\usepackage[warn]{mathtext}
\usepackage{amsmath}
\usepackage{fancyref}

\usepackage[figuresright]{rotating}
\usepackage{amsfonts,amssymb,amsmath}
\usepackage{boxedminipage}
\usepackage{makeidx}
\usepackage{url}

\usepackage[T2A]{fontenc}
\usepackage{latexsym}

\RequirePackage{snapshot}

\usepackage[english]{babel}
\usepackage{graphics}
\usepackage{epsfig}

\usepackage[table]{xcolor}
\usepackage{color,soul}
\usepackage{todonotes}
\usepackage{ulem}

\usepackage{cite}
\graphicspath{{img/}} 

\title
{Obtaining tensor-polarized nuclei with spin $S\ge1$ via spin filtering upon nuclear beam transmission through an unpolarized target (spin dichroism effect)}

\author{Sergey Anishchenko, Vladimir Baryshevsky, Alexandra Gurinovich}

\begin{document}
	
\maketitle

\section*{\small{{Abstract}}}
%
Upon transmission through an unpolarized target initially unpolarized nuclei with spin $S\ge1$ acquire tensor polarization due to spin dichroism effect.
%

%
For nuclei 
passing several nuclear lengths in the target, the acquired component $p_{zz}$ can reach substantial values of~$\sim0.2$--$0.8$.
%
%
The beam obtained in this way can be effectively used to study reactions with tensor-polarized nuclei.




\section{Introduction}

The study of atomic nuclei collisions at high-energies is currently one of the most vital tool of experimental and theoretical research in relativistic nuclear physics.
Transmission experiments are a method for determining nuclear shapes and matter densities, which is essential for uncovering exotic isotope structures and validating nuclear interaction models.
Next-generation high-energy nuclear physics experiments are shifting toward electron-ion colliders, where, in close analogy to nucleus-nucleus collisions, high-energy electrons probe target ions via deep inelastic scattering, enabling the mapping of even the spatial distribution of the quarks and gluons inside them.
All these methods are rapidly evolving at many laboratories, for example, at CERN~\cite{2019CMS,2022ALICE,2023ATLAS}, Brookhaven National Laboratory~\cite{2024STAR,2026STAR},  Nuclotron\,M--NICA complex (JINR, Dubna)~\cite{Malakhov2007,Baryshevsky2026} and 
Jefferson Lab~{\cite{JeffLab1,JeffLab2,JeffLab3}}.
%


%
%
It was demonstrated for the first time in papers~\cite{VG1992,VG1993}  that
when a beam of particles (nuclei) with spin $S\ge1$ passes through an unpolarized target  the quasi-optical phenomenon of birefringence occurs, leading to oscillations between vector and tensor polarizations and the spin dichroism effect.

%
%
%
The spin dichroism phenomenon  manifests via acquisition of tensor polarization by an initially unpolarized beam of nuclei  with spin $S\ge1$ passing through an unpolarized target.   
The effect  is due to the dependence of the number of particles passing through the target on the spin projection $m$ onto the direction of motion 
(where $m$ takes $2S+1$ values ranging from $-S$ to $S$), because
each value of~$|m|$ is characterized by a specific scattering cross section and absorption coefficient.
In other words, spin filtering occurs:  particles with those spin projections $m$, for which absorption in the matter is minimal, predominate in the transmitted beam.
Consequently, the spin dichroism effect can be used to produce tensor-polarized particle beams.
%


The dependence of the cross sections on $|m|$ arises due to the intrinsic anisotropy that particles (nuclei) with spin $S\ge1$ can possess. 
For particles with spin $S<1$, such anisotropy and the spin dichroism effect are absent in unpolarized matter, when $P$- and $T$-violating interactions are neglected.

%
%
%
To date, spin dichroism has been observed only for deuterons and in only two experiments: at the University of Cologne at a deuteron energy of 20~MeV~\cite{VG2005,VG2006,Seyfarth2010} and at the Nuclotron at JINR at a momentum of 5~GeV/c~\cite{Azhgirej2008,Azhgirej2009}.
%
%
%
Meanwhile, there are many nuclei with spin $S\ge1$~\cite{1975BohrMottelson,1980Ring,2021Obertelli} for which spin dichroism in an unpolarized target can also take place.
%
%
%
{The study of nuclear deformations in high-energy collisions represents a frontier area of both experimental and theoretical research in relativistic nuclear physics~\cite{2019CMS,2022ALICE,2023ATLAS,2024STAR,2025Otsuka}.}
%
%
%
Therefore, it seems appropriate to consider the acquisition of tensor polarization by a beam of deformed nuclei with spin $S\ge1$ possessing a significantly larger intrinsic anisotropy than the deuteron, after passing through an unpolarized target.


%

The present paper is dedicated to evaluating the degree of tensor polarization of various deformed nuclei with spin $S \ge 1$.
The paper is organized as follows. First, a semiclassical theory of spin dichroism for deformed nuclei is developed.
Then, using this theory, the tensor polarization components are evaluated for two specific cases: the passage of initially unpolarized beams of $^{21}\text{Ne}$ and $^{25}\text{Mg}$ nuclei through a carbon target.

\section{Spin dichroism of nuclei}
%
%
Let us consider the passage of a beam of deformed nuclei through a target of thickness $l$ with a scatterer density~$n$.
%
%
Since various papers use different notation for the same quantities, and because normalization conventions vary, we will explicitly point out how the formulas can be transformed between different notations throughout the text.

%
Attenuation of the transmitted beam during the passage through the target depends on the spin state of the particles (the value of the spin projection $m$ along the direction of motion) and is described as follows
\footnote{In~\cite{Plis1972} symbol $N(m)$ denotes the fraction of particles with spin projection $m$ rather than the number of particles with this projection.}:
\begin{equation}
	\label{eq:Nm}
	N^{(m)}(l)=\frac{N_0e^{-n\sigma^{(m)}l}}{2S+1}.
\end{equation}
Here $N_0$ is the initial number of particles, $\sigma^{(m)}$ is the total scattering cross-section for particles with spin projection~$m$, which takes $2S+1$ values ranging from $-S$ to $S$.

%

Assume the beam is initially unpolarized, i.e., it has an equal population in each possible spin state, so that $N^{(m)}(0) = N_0/(2S+1)$ for any value of~$m$.
The spin state of the beam propagating through the target is mixed and described by a diagonal density matrix:
\begin{equation}
	\label{eq:rho}
\hat\rho=\frac{1}{\sum_{m=-S}^{m=S}N^{(m)}(0)}\text{diag}\big(N^{(-S)}(l),...,N^{(S)}(l)\big),
\end{equation}
%
%
the sum of whose diagonal elements at $l=0$ is normalized to unity
$$\text{Sp}{\hat\rho}=1.$$

%
The density matrix~$\hat\rho$ is a convenient tool for describing the tensor polarization acquired by a nuclear beam.
For a quantitative description, let us consider a set of quadrupolarization  operators (rank 2 operators) as follows:
\begin{equation}
	\hat P_{ik}=\frac{3}{2S(2S-1)}\Big(\hat S_i\hat S_k+\hat S_k\hat S_i-\frac{2}{3}\hat S^2\delta_{ik}\Big),
\end{equation}
the expectation value of each of which is given by the expression as follows:
\begin{equation}
\langle\hat P_{ik}\rangle=\frac{\text{tr}(\hat{\rho}\hat{P}_{ik})}{\text{tr}(\hat{\rho})}.
\label{eq:Pik}
\end{equation}
%
Here, the indices $i$ and $k$ run over the values $1, 2, 3$, which correspond to the three Cartesian coordinates.

%
According to~\eqref{eq:rho} and~\eqref{eq:Pik}, for a  beam of nuclei with spin~$S$ moving along the $z$-axis, the expectation value $\langle\hat P_{zz}\rangle \equiv p_{zz}^{(S)}$ is calculated using the following
formula~\footnote{In~\cite{VG1993} quantity $p_{zz}^{{(S)}}$ is denoted as  $Q_{zz}$. Ohlsen~(see equation (2.28) in \cite{Ohlsen1972}) uses parameter $f_2$ to denote the 'degree of orientation'   along with 
 $p_{zz} \equiv p_{zz}^{{(S)}}$ differing only by a factor: $f_2=\frac{(2S-1)p_{zz}}{3S}$.}:
%
%
\begin{equation}
	\label{eq:Pzz}
p_{zz}^{{(S)}}=\frac{\sum_{m=-S}^{m=S}(m^2-S(S+1)/3)N^{(m)}(l)}{S(2S-1)\sum_{m=-S}^{m=S}N^{(m)}(l)}.
\end{equation}

%
In case of particles with spin $S=1$, $S=3/2$ and $S=5/2$
the tensor polarization component $p_{zz}^{{(S)}}$ equals respectively as follows:
\begin{equation}
	\label{eq:Pd}
	p_{zz}^{(1)}=1-\frac{3N^{(0)}(l)}{N^{(0)}(l)+2N^{(1)}(l)}=\frac{2(e^{n(\sigma^{(0)}-\sigma^{(1)})l}-1)}{2e^{n(\sigma^{(0)}-\sigma^{(1)})l}+1},
\end{equation}
\begin{equation}
	\label{eq:Pne}
	p_{zz}^{(3/2)}=\frac{N^{(3/2)}(l)-N^{(1/2)}(l)}{N^{(3/2)}(l)+N^{(1/2)}(l)}=\tanh\frac{n(\sigma^{(1/2)}-\sigma^{(3/2)})l}{2}
\end{equation}
and
\begin{equation}
	\label{eq:Pmg}
	\begin{split}
	&p_{zz}^{(5/2)}=\frac{N^{(5/2)}(l)-N^{(3/2)}(l)/5-4N^{(1/2)}(l)/5}{N^{(5/2)}(l)+N^{(3/2)}(l)+N^{(1/2)}(l)}\\
	&=\frac{e^{-n\sigma^{(5/2)}l}-e^{-n\sigma^{(3/2)}l}/5-4e^{-n\sigma^{(1/2)}l}/5}{e^{-n\sigma^{(5/2)}l}+e^{-n\sigma^{(3/2)}l}+e^{-n\sigma^{(1/2)}l}}\,.\\
	\end{split}
\end{equation}
%
Suppose the differences between the scattering cross sections in states with 
different spin projections~$m$ are much smaller than $1/nl$. Then, formulas 
\eqref{eq:Pd}--\eqref{eq:Pmg} reduce to the following approximate form:
\begin{equation}
	\label{eq:PdL}
	p_{zz}^{(1)}\approx\frac{2n(\sigma^{(0)}-\sigma^{(1)})l}{3},
\end{equation}
\begin{equation}
	\label{eq:PneL}
	p_{zz}^{(3/2)}\approx\frac{n(\sigma^{(1/2)}-\sigma^{(3/2)})l}{2}
\end{equation}
and
\begin{equation}
	\label{eq:PmgL}
	p_{zz}^{(5/2)}\approx n(4\sigma^{(1/2)}/5+\sigma^{(3/2)}/5-\sigma^{(5/2)})l.
\end{equation}

%
Note that we should distinguish transmitted and detected beams. Formula~\eqref{eq:Nm} describes the attenuation of a particle beam that 
has passed through the target and experienced neither inelastic nor 
incoherent elastic scattering by individual scatterers.
%
%
%
Meanwhile, when the target thickness is a few nuclear lengths (mean free paths), the average 
number of nuclear collisions turns out to be close to unity.
%
%
If the detection system cannot distinguish incoherently scattered nuclei, 
these nuclei also contribute to the measured characteristics of the 
transmitted beam, including the tensor polarization components. 
%

%
Multiple scattering theory should be used for the rigorous description of the dichroism phenomenon. However, in the simplest case, estimations can be performed by using the reaction cross-section instead of the total cross-section, which accounts for all elastic and inelastic reaction processes.
In this case, the attenuation of the detected beam intensity 
is determined by the absorption cross sections~$\sigma_a^{(m)}$~\cite{Baryshevsky2012} 
rather than the total cross sections~$\sigma^{(m)}$.
%
%
Therefore, to evaluate component $p_{zz}^{(S)}$
in \eqref{eq:Pzz}--\eqref{eq:PmgL} one should replace $\sigma^{(m)}$  with $\sigma_a^{(m)}$.


To estimate the cross sections~$\sigma_a^{(m)}$, let us consider the collision 
of a nucleus with mass number~$A$ and a target composed of identical nuclei with mass number~$A_t$. 
Suppose that the incident deformed 
nuclei are prolate or oblate ellipsoids of revolution described by 
the small deformation parameter~$\beta_2$~\cite{1975BohrMottelson,1980Ring,2021Obertelli}, 
while the target nuclei are considered spherical. 
The value of~$\beta_2$ is related to the semi-axes $a$ and $b$ of the ellipsoid as follows:
\begin{equation}
	\begin{split}
		& a=R_0\Big(1+\sqrt{\frac{5}{4\pi}}\beta_2\Big),\\
		& b=R_0\Big(1-\sqrt{\frac{5}{16\pi}}\beta_2\Big),\\
	\end{split}
\end{equation}
%
where $R_0 = r_0 A^{1/3}$~fm is the nuclear radius (root-mean-square (rms) radius of a nucleus), $A$ is the mass number, 
and $r_0 \approx 1.25$~fm. 
The unpolarized target nuclei are assumed 
to be ideal spheres with radii $r_t = r_0 A_t^{1/3}$.

%
In the case of completely opaque nuclei, the absorption cross section~$\sigma_a^{(m)}$ 
is equal to the geometric shadow area of the collision~\cite{Valanten1986}. 
For spherical nuclei ($a=b$), whose projections onto the collision plane are circles, 
this area equals~$\pi(a+r_t)^2$.
%
%
%
Since the incident nuclei are ellipsoids of revolution, their projections 
onto the collision plane are ellipses with semi-axes $a_p=\sqrt{a^2\sin^2\theta+b^2\cos^2\theta}$
and $b$, respectively, where $\theta$ is the angle between the direction of motion 
and the spin of the incident particle.
%
%
In this case, the collision shadow determined by the geometric touch point  
of the colliding nuclei is approximately elliptical, with semi-axes 
$a_p + r_t$ and $b + r_t$. 
For small~$\beta_2$, the collision shadow area, 
which equals the absorption cross section, reads:
\begin{equation}
	\label{eq:S}
	\sigma_{a\theta}\approx\pi r_0^2(A^{1/3}+A_t^{1/3})^2+\frac{\beta_{2}r_0^2A^{1/3}\sqrt{5\pi}(A^{1/3}+A_t^{1/3})}{4}\cdot\Big(1-3\cos^2\theta\Big).
\end{equation}
%
%
Recalling that~$\theta$ is the angle between the particle spin direction and the $z$-axis, 
one can use the following quantum-mechanical estimate~\cite{Born1965}:
\begin{equation}
\cos^2\theta\approx\frac{m^2}{S(S+1)}.
\end{equation}
%
Therefore, formula \eqref{eq:S} reads:
\begin{equation}
	\label{eq:sigmam}
	\sigma_a^{(m)}\approx\pi r_0^2(A^{1/3}+A_t^{1/3})^2+\frac{\beta_{2}r_0^2A^{1/3}\sqrt{5\pi}(A^{1/3}+A_t^{1/3})}{4}\cdot\Big(1-\frac{3m^2}{S(S+1)}\Big).
\end{equation}
%
If the beam is unpolarized, the absorption cross section averaged over the spin states equals
\begin{equation}
	\sigma_a=\frac{1}{2S+1}\sum_{m=-S}^{m=S}\sigma_a^{(m)}.
	\label{eq:cross-section}
\end{equation}
%
%
%
For diffractive scattering by completely opaque target nuclei, the total 
cross sections~$\sigma^{(m)}$ are twice as large as the absorption 
cross sections~$\sigma_a^{(m)}$; in accordance with \eqref{eq:cross-section} the following is also valid $\sigma\approx2 \sigma_\text{a}$.

Thus, the difference between the absorption cross sections $\Delta\sigma_a^{(S)}$ expressed 
in terms of the quadrupole deformation parameter~$\beta_2$ reads for a particle with spin~$1$:
\begin{equation}
	\Delta\sigma_a^{(1)}=\sigma_{a}^{(0)}-\sigma_{a}^{(1)}=\frac{3\sqrt{5\pi}r_0^2A^{1/3}\beta_2(A^{1/3}+A_t^{1/3})}{8},
\end{equation}
and for a particle with spin~3/2:
\begin{equation}
	\Delta\sigma_a^{(3/2)}=\sigma_{a}^{(1/2)}-\sigma_{a}^{(3/2)}=\frac{2\sqrt{\pi}r_0^2A^{1/3}\beta_2(A^{1/3}+A_t^{1/3})}{\sqrt{5}}.
\end{equation}

\section{Nuclear beams with tensor polarization} \label{ch:heavy}

%
As mentioned above, the birefringence effect associated with particle motion 
through unpolarized matter is caused by the particle's internal anisotropy 
and described quantitatively by the deformation parameter~$\beta_2$.
For example, for $^{21}$Ne nucleus ($S=3/2$) and for $^{25}$Mg nuclei ($S=5/2$) the deformation parameter 
 $\beta_2$ equals $0.463$ and  $0.384$, respectively.
Many other nuclei also exhibit significant internal anisotropy.

Therefore, an initially unpolarized beam of nuclei with spin $S \ge 1$ 
passing through an unpolarized target acquires tensor polarization due to spin dichroism, whose value increases 
with the path length in the target. 
Simultaneously, the longer the path length in the target, the higher the beam attenuation.
If the beam passes 
	distance $l=4.6 \cdot L_\text{a}$ in the target, it is attenuated 100 times ($N/N_0=10^{-2}$) due to absorption,  while an attenuation 
	by a factor of $10^4$ corresponds to $l = 9.2 \cdot L_{\text{a}}$, 
	where $L_{\text{a}} = \frac{1}{n \sigma_{\text{a}}}$.

\begin{figure}[h]
	\begin{center}
		\resizebox{!}{80mm}{\includegraphics{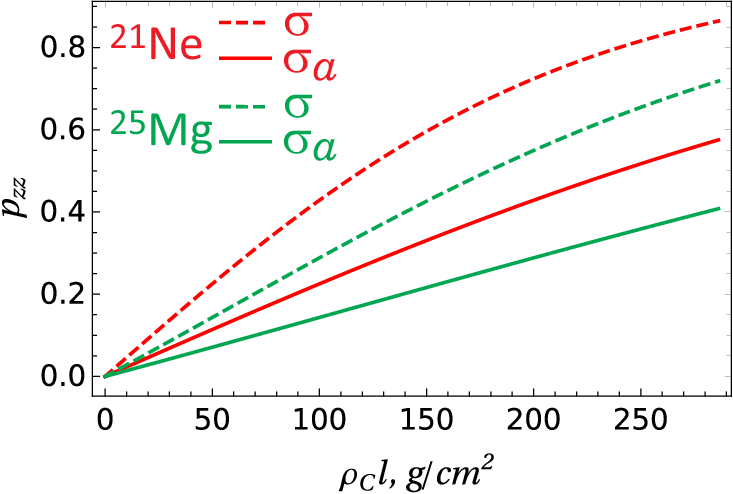}}
	\end{center}
%
	\caption{Dependence of the tensor polarization~$p_{zz}$ acquired by $^{21}$Ne (red lines) 
		and $^{25}$Mg (green line) nuclei passing through carbon targets. 
		Dashed lines correspond to nuclei that do not participate in incoherent scattering. Solid lines correspond to nuclei experienced either coherent or incoherent elastic scattering.}
	\label{fig:NeMg}
\end{figure}

\begin{table}[h]
	\setlength{\extrarowheight}{2pt}
	\renewcommand{\arraystretch}{1.6}
	%
	%
\caption{Calculated values of tensor polarization acquired by $^{21}$Ne and $^{25}$Mg nuclei in carbon targets due to coherent and incoherent scattering.}
	\label{tab:Ne-Mg}	
	\centering
	\begin{tabular}{|c|c|c|c|c|}
		\hline
			& \multicolumn{2}{c|}{$^{21}$Ne}& \multicolumn{2}{|c|}{$^{25}$Mg}\\
		\hline
		$\rho_C l$, g/cm$^2$ & $N/N_0$ & $p_{zz}$ & $N/N_0$ & $p_{zz}$ \\
		\hline
		65  & 0.097              & 0.15 & 0.12              & 0.09  \\
		\hline
		130 & $0.94\cdot10^{-2}$ & 0.29 & $1.5\cdot10^{-2}$ & 0.19\\
		\hline
		260 & $0.88\cdot10^{-4}$ & 0.54 & $2.3\cdot10^{-4}$ & 0.37\\
		\hline
	\end{tabular}
\end{table}
\begin{table}[h]
	\setlength{\extrarowheight}{2pt}
	\renewcommand{\arraystretch}{1.6}
	%
	%
\caption{Calculated values of tensor polarization acquired by $^{21}$Ne and $^{25}$Mg nuclei in carbon targets, that do not participate in incoherent scattering.}
\label{tab:Ne-MgCoherent}	
\centering
\begin{tabular}{|c|c|c|c|c|}
	\hline
	& \multicolumn{2}{c|}{$^{21}$Ne}& \multicolumn{2}{|c|}{$^{25}$Mg}\\
	\hline
	$\rho_C l$, g/cm$^2$ & $N/N_0$ & $p_{zz}$ & $N/N_0$ & $p_{zz}$ \\
	\hline
	65  & $0.94\cdot10^{-2}$ & 0.29 & $1.5\cdot10^{-2}$ & 0.19\\
	\hline
	130 & $0.88\cdot10^{-4}$ & 0.54 & $2.3\cdot10^{-4}$ & 0.37\\
	\hline
	260 & $0.77\cdot10^{-8}$ & 0.83 & $5.4\cdot10^{-8}$ & 0.67  \\
	\hline
\end{tabular}
\end{table}


%


%
According to the calculations, the tensor polarization of the $^{21}$Ne beam ($S=3/2$ and $\beta_2=0.463$)
in the carbon target reaches {$p_{zz}=0.29$} at $\rho_Cl = 130$~g/cm$^2$ 
and {$p_{zz}=0.54$} at $\rho_C l = 260$~g/cm$^2$ (see Table~\ref{tab:Ne-Mg} and the solid 
lines in Figure~\ref{fig:NeMg}).
Calculations performed for $^{25}$Mg nuclei ($S=5/2$ and $\beta_2=0.384$) 
passing through a carbon target yield similar results: tensor polarization values 
of $p_{zz} \sim 0.19$ at $\rho_C l = 130$~g/cm$^2$ and $p_{zz} = 0.37$ at $\rho_Cl = 260$~g/cm$^2$ 
can be obtained.

%
%
The solid curves in Figure~\ref{fig:NeMg} are obtained assuming that both 
participating and non-participating particles in incoherent elastic scattering 
are detected. 
Their loss in the target is determined by the absorption 
cross sections~$\sigma_\text{a}$.

%
%
If it is experimentally possible to extract particles that do not undergo 
incoherent scattering, the attenuation of the detected beam  should 
be determined by the total cross sections~$\sigma \approx 2\sigma_\text{a}$.

%

The dimensionless quantities $\sigma n l$, on which $p_{zz}^{(S)}$ depends, 
turn out to be twice as large as $\sigma_a n l$ for a fixed target thickness~$l$. 
Therefore, the degree of tensor polarization of the beam is also higher 
(see dashed lines in Figure~\ref{fig:NeMg}), however beam is attenuated more in this case. 
For instance, after passing through 
a target with thickness $\rho_Cl = 260$~g/cm$^2$, {the tensor 
	polarization component $p_{zz}$ of $^{21}$Ne nuclei reaches~$0.83$.}

%

Thus, the estimates performed in this paper demonstrate that the values of the 
$p_{zz}$ tensor polarization component acquired by deformed $^{21}$Ne and $^{25}$Mg 
nuclei in a carbon target due to spin dichroism effect are quite significant. 
This fact makes it possible to use the beams produced in this way for studying nuclear reactions with tensor-polarized nuclei.

%

The high degree of tensor polarization acquired by nuclei with spin $S \ge 1$ due to spin dichroism at low energies 
	makes their further acceleration and storage in the NICA ring a perfect tool for  
 investigating scattering processes and reactions 
	involving tensor-polarized nuclear beams~\cite{SPIN2025}.

%


\section{Conclusion}	
%
%
In a carbon target with a thickness of several nuclear absorption lengths, 
an initially unpolarized beam of deformed $^{21}$Ne and $^{25}$Mg nuclei 
moving along the $z$-axis acquires a tensor polarization component $p_{zz}$ 
as high as $\sim 0.2$--$0.8$, that is accompanied by simultaneous beam attenuation.
%
%
%
 The $p_{zz}$ value of the nuclear beam passing through the target is strongly dependent on  the possibility of extracting a part of the beam  that does not undergo incoherent scattering on the target nuclei.
%
%
Large $p_{zz}$ values make the beams obtained via spin filtering suitable 
for the investigation of reactions with tensor-polarized nuclei.
%
%
%
%
Note that implementing experiments to obtain tensor-polarized nuclear beams 
requires a detailed analysis of higher-spin particle polarimetry and the 
extraction of the transmitted beam from background particles.

\end{document}